\documentclass[twocolumn,prl,showpacs,floatfix]{revtex4}
\usepackage{graphicx}
\usepackage{epsfig}
\usepackage{rotating}
\usepackage{amsmath}
\usepackage{amsfonts}
\usepackage{amssymb}
\usepackage{enumerate}
\usepackage{longtable}
\usepackage{dcolumn}
\usepackage{bm}

\begin{document}

\title{Valence Bond Glass Phase in the Diluted Kagome Antiferromagnets}

\author{R. R. P. Singh}
\affiliation{University of California Davis, CA 95616, USA}

\date{\rm\today}

\begin{abstract}
We present a theory for site dilution in the Valence Bond 
Crystal Phase of the Kagome Lattice Heisenberg Model.
The presence of an empty site leads
to strong singlet bonds across the impurity. It also
creates a free spin, which delocalizes inside the unit cell.
Finite concentration
of quenched impurities leads to a Valence Bond Glass phase. 
This phase has short-range Valence Bond order,
no spin-gap, large spin-susceptibilities, linear specific heat due
to two-level systems,
as well as singlet and triplet excitations that decompose
into kink-anikink pairs delocalized over a few lattice constants.
It provides a framework for a comprehensive understanding 
of thermodynamic, neutron, and Raman measurements in the Herbertsmithite 
material
ZnCu$_3$(OH)$_6$Cl$_2$, including recently reported $H/T$ and $\omega/T$
scaling.
\end{abstract}


\maketitle
The Kagome Lattice Heisenberg Model (KLHM)
is one of the most studied realistic quantum spin-models where 
Resonating Valence Bond (RVB) physics, first proposed by Anderson\cite{pwa} more than twenty five years ago,
may be realized.
In this model, extensive degeneracy at the classical 
level and in the space of Valence Bond configurations leads to the possibility of quantum
spin liquids as well as many competing or coexisting orders.\cite{elser,mila,misguich,palee,mpaf,VBC,
ed,dmrg,singh-huse,ybkim,ms07,laeuchli,vidal,tchernyshyov,poilblanc}.

Experimental studies of the Herbertsmithite materials ZnCu$_3$(OH)$_6$Cl$_2$,
containing structurally perfect Kagome planes,\cite{shores,helton,keren,shlee,olariu,imai,devries1} 
have raised hope of realizing RVB physics in a real material.
Ideally, these materials have
a pyrochlore structure, where spin-half copper atoms form Kagome layers, which are
separated by non-magnetic zinc containing triangular layers. This ideal case is now
known to be not true as substitution of some fraction of zinc and copper sites leads
to extra isolated spins in the zinc planes and site-dilution in the copper planes.

Here, we develop a theory for site dilution in KLHM  in context of Dimer
series expansions. \cite{singh-huse,singh-huse2}
We find that dilution leads to a Valence Bond Glass phase, which has short-range
Valence Bond order, no spin-gap, large spin-susceptibility, linear
specific heat due to two-level systems as well as singlet and triplet 
excitations that decompose into kink-antikink pairs with energies
spread upto approximately $2.5 J$. 
Unlike the VBC phase, which is stabilized by higher order
quantum fluctuations, with energy scales less than $0.01$ J per site (energy scales that differentiate different VBC
phases are only $0.001$ J per site), the VBG phase is stabilized
by second order perturbation theory with an energy scale of order $0.1$ J per site. 
This VBG phase provides a novel framework for a comprehensive 
understanding of the Herbertsmithite materials.

\begin{figure}
\begin{center}
\includegraphics[width=0.8\columnwidth,clip,angle=0]{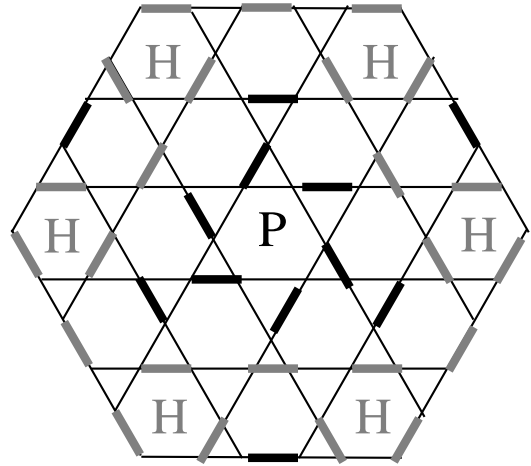}
\caption{\label{fig:Fig1} Proposed Valence Bond Phase of the Kagome Lattice
Heisenberg Model consists of a Honeycomb Lattice of resonating hexagons (H),
where six hexagons surround a pinwheel (P). The pinwheels are isolated from empty triangles
leading to substantially reduced quantum fluctuations and doubly degenerate dimer configurations.
}
\end{center}
\end{figure}

A single impurity can be accomodated in the Valence Bond Crystal phase
in several inequivalent positions of the 36-site unit cell
with nearly equal energy. The addition of such an impurity leads
to strong singlet bonds across from it, in agreement
with earlier studies of Dommange et al.\cite{dommange} 
However, contrary to the work of Dommange et al, we find that the impurity also
creates a free spin, which delocalizes inside the unit cell.
The Herbertsmithite materials have a large impurity
concentration of order six percent, which would imply on average 2 randomly
placed impurities per unit cell. 
Combined with the singlet bonds across from the impurities, it also means
30 percent of spins are inactive. 
At this high impurity concentration no semblence of long range VBC order would remain.
Instead, one obtains a randomly pinned Valence Bond Glass phase with only short
range Valence Bond order.
We will show that the VBG description leads to a consistent picture for
thermodynamics, neutron and Raman scattering experiments in the Herbertsmithite materials.

The dimer expansions provide a hirearchy of energy scales for different order quantum
fluctuations in the KLHM ( we set $J=1$).\cite{singh-huse} A dimer configuration has energy $-0.375$. After
including second order quantum fluctuations, the energy of  each dimer configuration is lowered
to $-0.421875$. We emphasize that at this level all dimer configurations lead to the same energy.
On the other hand the exact ground state energy of KLHM is known from a variety of numerical studies
to be $-0.433$. Thus the entire set of higher order quantum fluctuations beyond second order change the energy of
the system by only another $-0.011$. The first selection in the space of dimers comes from the
occurance of resonating hexagons.\cite{VBC} In 3rd order, having a maximally
allowed number of such hexagons lowers the energy of the state to $-0.42578125$. These hexagons
also contribute a fraction of the fourth order perturbation theory, that brings the energy to approximately
$-0.431$. The favored VBC configuration, which is selected in fourth order perturbation theory is shown in Fig.~1.
We should note that the same VBC configuration has been obtained variationally in a completely
unbiased tensor-network formalism by Evenbly and Vidal.\cite{vidal}

The addition of holes lifts the degeneracy of the dimer configurations 
already in second order perturbation theory. For a finite concentration of holes, not having a strong
bond in a triangle with a missing spin raises the energy of the dimer configuration by about $0.05$
if one bond opposite the missing spin is absent and $0.1$ if both bonds opposite the missing spin are absent. Given this
relatively large energy scale, it is fair to assume that these impurities freeze all 
quantum fluctuations  except those involving (i) a pair of neighboring dimers (second order perturbation), 
(ii) resonating hexagons and (iii) pinwheels and analogous structures. 
Because the pinwheels have a doubly degenerate dimer configuration, they play a vital role in the VBG phase at
low impurity concentration.

\begin{figure}
\begin{center}
\includegraphics[width=0.8\columnwidth,clip,angle=0]{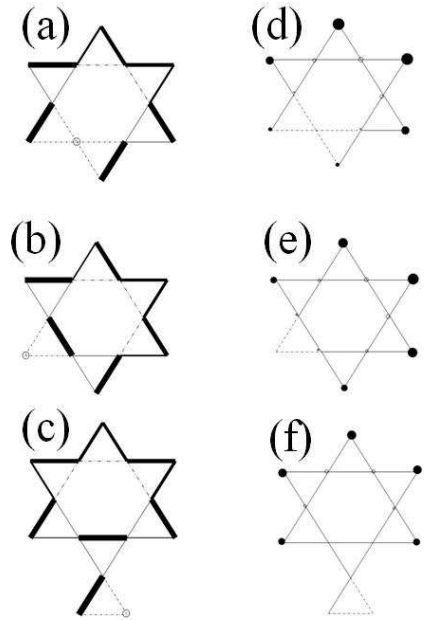}
\caption{\label{fig:Fig2} 
Dimer configurations around the impurity, when the impurity is placed
(i) in the inner hexagon, (ii) on the outer vertices, and (iii) in the
bond neighboring the pinwheel. Fig (a) through (c) show the dimerization
pattern, with the thickness of the lines showing the strength of the singlet
bonds. Note that the other side of the impurity in (b) and (c) is 
a non-fluctuating singlet bond. 
Fig (d) through (f) show the distribution of free spins. The filled 
circles show the positive spin expectation values, whereas empty circles 
show negative values. The size of the circle represents the size of the
spin.
}
\end{center}
\end{figure}

We first ask the question, how would the crystal respond to a single impurity? Given the many
inequivalent sites, the crystal is free to adjust itself to place the impurity at any site.
Second order perturbation theory tells us that the impurity will prefer to go to one of the
bonds that do not touch any empty triangles. These are shown as the dark bonds in Fig.~1.
They are the same bonds that lead to heavy (non-mobile) triplets in the VBC phase.\cite{singh-huse2}
To a high numerical approximation, there are three inequivalent sites where the missing spin
can go. These are (i) the inside hexagon of the pinwheel, (ii) The outside vertices of
the pinwheel and (iii) on one of the dark bonds that are not part of the pinwheel.
The structures are all bounded
from outside by dimerized triangles, which are in their ground state. Hence, it suffices to
diagonalize the pinwheels only, plus the extra triangle in case (iii).

The ground state bond energies and spin configurations for the three cases 
are shown in Fig.~2.
Fig. 2(a) through (c) show the pattern of dimerization around the impurity. 
Fig. 2(d) through (f) show how the free spin is delocalized over the pinwheel.
In all cases, the energy cost of removing a spin is not $0.75$ as for an isolated
dimer but rather approximately $0.25$. The system chooses to put a strong singlet
bond across the impurity, which gets back the lost $0.75$ of energy. Instead
an antikink develops in the pinwheel, which is known to have a minimum energy of approximately
$0.25$. 
The actual energy cost in the three examples is found by exact diagonalization to be approximately
0.2764, 0.2632 and 0.2468 in cases (i), (ii) and (iii) respectively. 
The singlet bond across the impurity has strength $-0.725$, $-0.716$ 
and $-0.731$ respectively. The bond patterns
have a clear resemblence to the study of Dommange et al.\cite{dommange} 

In contrast to the
dark bonds, if the impurity is placed on the grey bonds, the resulting free spin can delocalize
through the network of grey bonds. It cannot delocalize too far because of the 
confining dimerizing field, but
the dimerization between the weak and strong bonds connecting the empty triangles is relatively
weak. This implies that the confining potential is weak and the spin can delocalize over several unit cells.

With a finite concentration of randomly placed impurities, there will be
lots of free spins created.
When two of these spins meet they would bind into a singlet.\cite{tchernyshyov}
The mobile spins
can also wander into the pinwheel regions and form singlets with spins there.
The strength of the pairing will depend on the ability of the spins to be in
overlapping regions and thus would decrease rapidly with separation of impurities.
In a thermodynamic system of randomly placed impurities, most of the spins would be paired into singlets at T near zero. 
Only very rare regions with isolated impurities will have a free spin. 
This implies a zero spin-gap for the system.
At infinitesimally small impurity concentration there would be a 1/T susceptibility with
a strength proportional to the impurity concentration. But, at any significant impurity concentration
this strength would be reduced. 
We show in appendix that the assumption that these spins go into a random 
singlet phase\cite{dsf} with a power-law distribution of exchange constants, allows us 
to reproduce the $H/T$ and $\omega/T$ scaling reported recently.\cite{helton10}

The glassy system has many potential two-level system type local excitations 
created by the impurities. For very dilute impurities, the
pin-wheels would serve as two level systems. The random environment would produce a weak random exchange
field on the pinwheels giving rise to a small splitting, between its two ground states. 
In a higher concentration range, as relevant
for the Herbertsmithite materials, very few intact pinwheels are likely to remain.
Instead, at larger impurity concentration,
the two-level systems would be a network of corner 
sharing triangles, where the Valence Bonds can switch between 
alternate dimer configurations, with very little energy cost. 
These would resemble Delta chains or Husimi trees,\cite{kubo,shastry,tchernyshyov}
a pinwheel being just one special case.
If we assume one
two level system per 36-sites and further assume that their splitting can range upto $10$ percent of $J$ (approximately 20 K)
that would lead to a density of states of $1/(720 K)$ per copper atom. This will give rise to a linear
term in specific heat of approximately $20 mJ/mol K^2$ per copper atom. 
This is in the right ballpark of the experimental
observations on the Herbertsmithite materials
when the spin degrees of freedom associated with free spins
in the zinc planes, are supressed by a magnetic field.\cite{helton-pc}

The very low energy magnetic response will depend on the DM interactions,\cite{rigol07,sindzingre07,cepas} which are
not considered here. For example, the NMR
spectra has already been addressed by Rousochatzakis et al in context of frozen dimers
and it requires non-zero DM interactions
in addition to impurities.\cite{mendels,mila10} The VBG phase also provides a natural explanation for
the observed Neutron spectra in the materials. 
One of the key observations of
recent powder diffraction neutron measurement of the materials by deVries et al\cite{devries} is that the
spectra is spread over a large frequency range and its angle averaged behavior is very close to that
of isolated dimers. 
This is exactly what one expects for pinwheels or Delta chains, as we show 
below.

\begin{figure}
\begin{center}
\includegraphics[width=0.9\columnwidth,clip,angle=0]{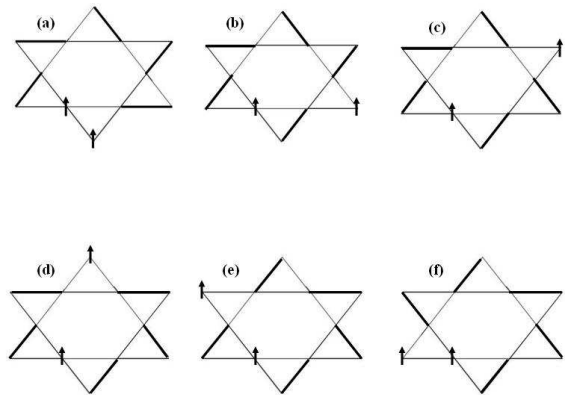}
\caption{\label{fig:Fig3} Kink-antikink or 2-spinon states for the pinwheel.
A triplet excitation is created by breaking a singlet bond in the ground state.
The Kink spinon lies on the inner hexagon, while the more mobile antikink
spinon can move on the outside vertices of the pinwheel through states (a) through
(f) shown in the figure. 
}
\end{center}
\end{figure}

Because the ground state of a pinwheel
is fully dimerized, the equal-time correlation function is strictly nearest-neighbor only,
that is, it is that of isolated dimers. However,
the triplet excitations in these systems decompose into a kink-antikink pair
as shown in Fig.~3. Hence the dynamic correlations are not
strictly nearest-neighbor only and unlike a single dimer they extend over a 
wide energy range.
It is known\cite{kubo,shastry} that
the kinks are gapless and anti-kink energy can be approximated by
$$\epsilon(k)=5/4-\cos{k},$$
so that the kink antikink pair energy ranges approximately from about $1/4$ to $9/4$. 
Indeed, we have found by exact diagonalization that $98$ percent of the spectral weight of the pinwheel is
spread over the energy $0.26$ to $2.50$.
Also, as shown in Fig.~4, the angle-integrated spectra at every
energy has q dependence that is very close to that of isolated dimers. This shows that
Delta chains have spectra spread over a wide energy range yet have angle integrated spectral
weight essentially that of isolated dimers, in agreement with experiments.

The presence of dilution also provides a simple explanation for the absence
of large peaks at low energy in the Raman spectra\cite{lemmens} that were found
in exact diagonalization studies of finite clusters.\cite{laeuchli} The pinning of dimers
destroys the low energy singlets, by making any long-range rearrangement of 
dimers energetically costly. Light scattering should produce
two antiparallel triplet excitations on neighboring dimers. Two
of these spins can combine into a singlet to leave a kink-antikink pair or two anti-kinks one of which 
is trapped in a triangle with a valence bond. In either case, this leads to
energy spread over the same range as the triplets, i.e., roughly upto $2.5 J$. 
Indeed, the exact diagonalization results of Laeuchli and Lhuillier\cite{laeuchli} show that once
the low energy peaks are removed from the spectra the frequency dependence of the neutron and Raman scattering profiles
are very similar. It supports the idea of two weakly interacting spinons which can form singlets or triplets. 
Hao and Tchernyshyov have shown\cite{tchernyshyov} that these
excitations have fermionic statistics. 
The minimum excitation energy is reduced when one doesnt have perfect pin-wheel structures
and the absence of spin-gap means that Raman scattering should also extend down to zero energy.
These features are, indeed, consistent with the 
experiments.\cite{lemmens}

\begin{figure}
\begin{center}
\includegraphics[width=0.8\columnwidth,clip,angle=270]{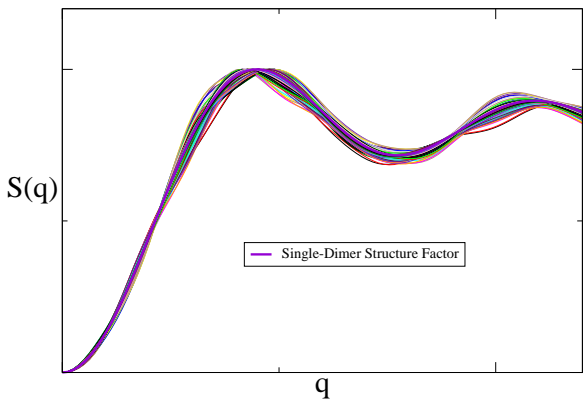}
\caption{\label{fig:Fig4} Exact diagonalization results for Angle averaged dynamic structure factor for a 
pinwheel, at different frequencies, scaled to 
have the same maximum, 
compared with results of a single dimer.}
\end{center}
\end{figure}

Since the disordered glass phase breaks no symmetry, there should be no phase
transition in the system and only a gradual crossover
from the high temperature paramagnet to the VBG phase. The temperature scale for the
freezing is set by $0.1 J$,
but some strong bonds may freeze at temperatures
approaching $J$ giving one a broad crossover region. Indeed the neutron
scattering measurements suggest that Dimer correlations persist beyond $T=J/2$.
On the other hand the Raman spectra show a clear development of a 
quasi-elastic peak at low energies as the temperature is raised from
$5K$ to $295 K$.\cite{lemmens} This is suggestive of a gradual melting
of the Valence Bond Glass into a classical Valence Bond Liquid as the 
frozen dimers are freed up, giving rise to lots of low energy
overdamped singlet excitations. 
Details of the Raman spectra, including dependence on polarization\cite{cepas2} deserve further theoretical attention.

In conclusion, we have developed a picture for the Valence Bond Glass phase when the 
Kagome Lattice Heisenberg Model is randomly diluted. We have argued that this
phase has no spin-gap, and supports local two-level system excitations as well as singlet and triplet
excitations that decompose into kink-antikink pairs that delocalize over a
local network of corner sharing triangles.
This phase provides a consistent picture of thermodynamic, neutron and Raman measurements
in the Herbertsmithite materials. Significantly reduced concentration of impurities can help provide
clearer signatures for the Valence Bond Crystal phase in the Kagome Lattice Heisenberg Model.

\begin{acknowledgements}
We would like to thank Fabrice Bert, Mark de Vries, Joel Helton, 
Michael Hermele,
Andreas Laeuchli, Peter Lemmens, Frederic Mila, 
and Oleg Tchernyshyov for many valuable discussions and to Phillipe Mendels for an invitation to 
Paris, where many of these ideas were crystalized.
\end{acknowledgements}

\section{Appendix}
In this appendix, we address the recetly reported Neutron scattering data on the
Herbersmithite materials by Helton et al.\cite{helton10}
We make the assumptions that the large number of free spins created by
the impurities go into a random singlet phase, where the spins are coupled by
a renormalized distribution of exchange constants, which has a power-law behavior
$P(J)=C J^{-\alpha}$.
With this the magnetization of the system as a function of field is given by
\begin{equation}
M(H,T)/T^{1-\alpha}=\int dx x^{-\alpha} f_1(y,x).
\end{equation}
with $y=gh/T$ with $h=\mu_B H$, and
\begin{equation}
f_1(y,x)={2 \sinh{y}\over e^x+1+2 \cosh{y}}
\end{equation}
The susceptibility in a field is given by
\begin{equation}
\chi(H,T)T^\alpha=\int dx x^{-\alpha} f_2(y,x)
\end{equation}
with
\begin{equation}
f_2(y,x)={2 \cosh{y} (e^x+1)+4 \over (e^x+1+2 \cosh{y})^2}
\end{equation}
Also, the imaginary part of the dynamic susceptibility for $\omega>0$
is given by
\begin{equation}
\chi^{\prime\prime}(\omega,T) T^\alpha=({T\over \omega})^\alpha
{e^{\omega/T}-1\over e^{\omega/T}+3}
\end{equation}

Motivated by the experimental observation, we will take $\alpha=0.66$.
We also take $g=2.2$. With these the plots for $M(H,T)$, $\chi(H,T)$
and $\chi^{\prime\prime}$ are shown in Figures. They appear remarkably
similar to the experiments.

\begin{figure}
\begin{center}
\includegraphics[width=0.7\columnwidth,clip,angle=270]{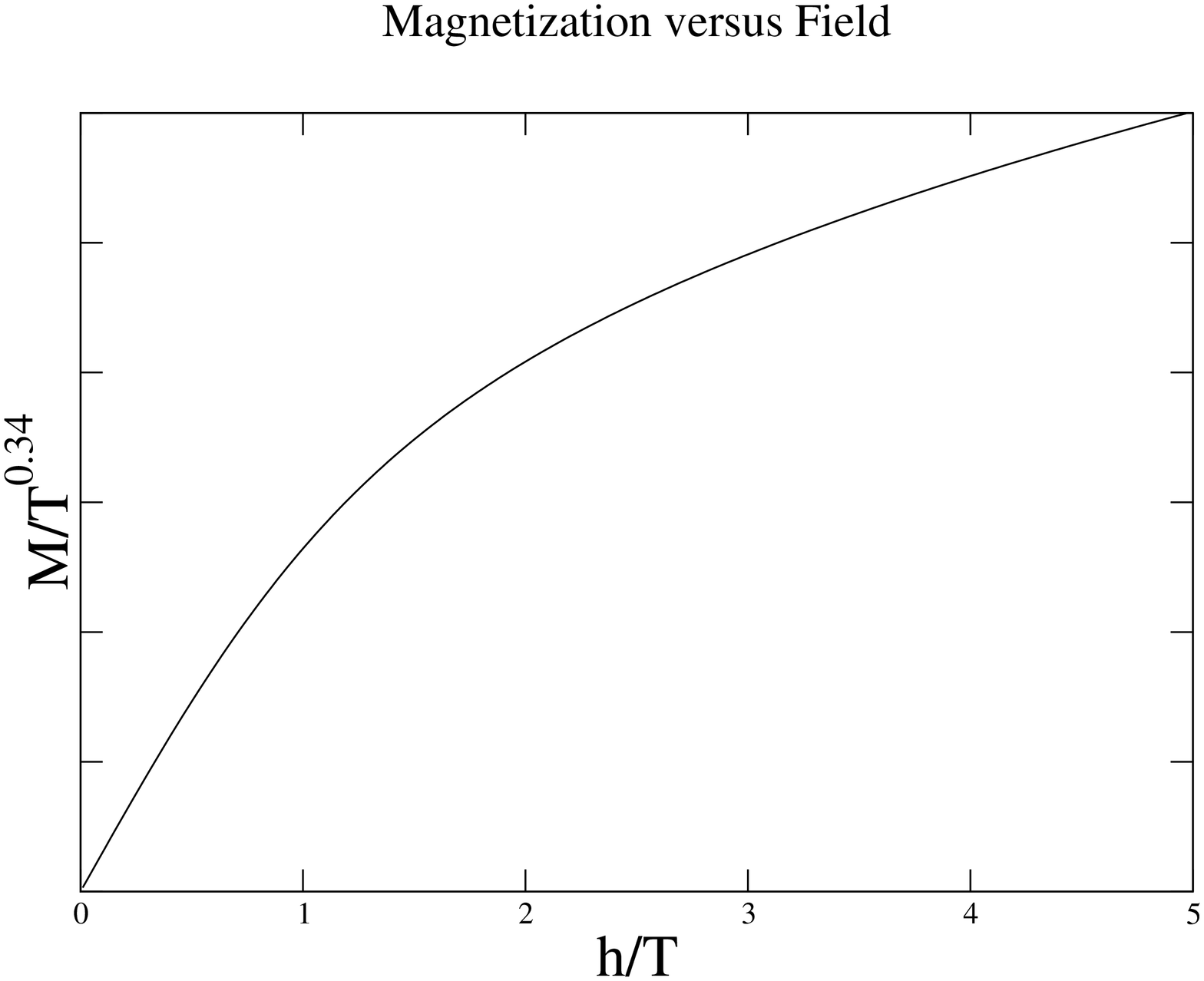}
\end{center}
\end{figure}

\begin{figure}
\begin{center}
\includegraphics[width=0.7\columnwidth,clip,angle=270]{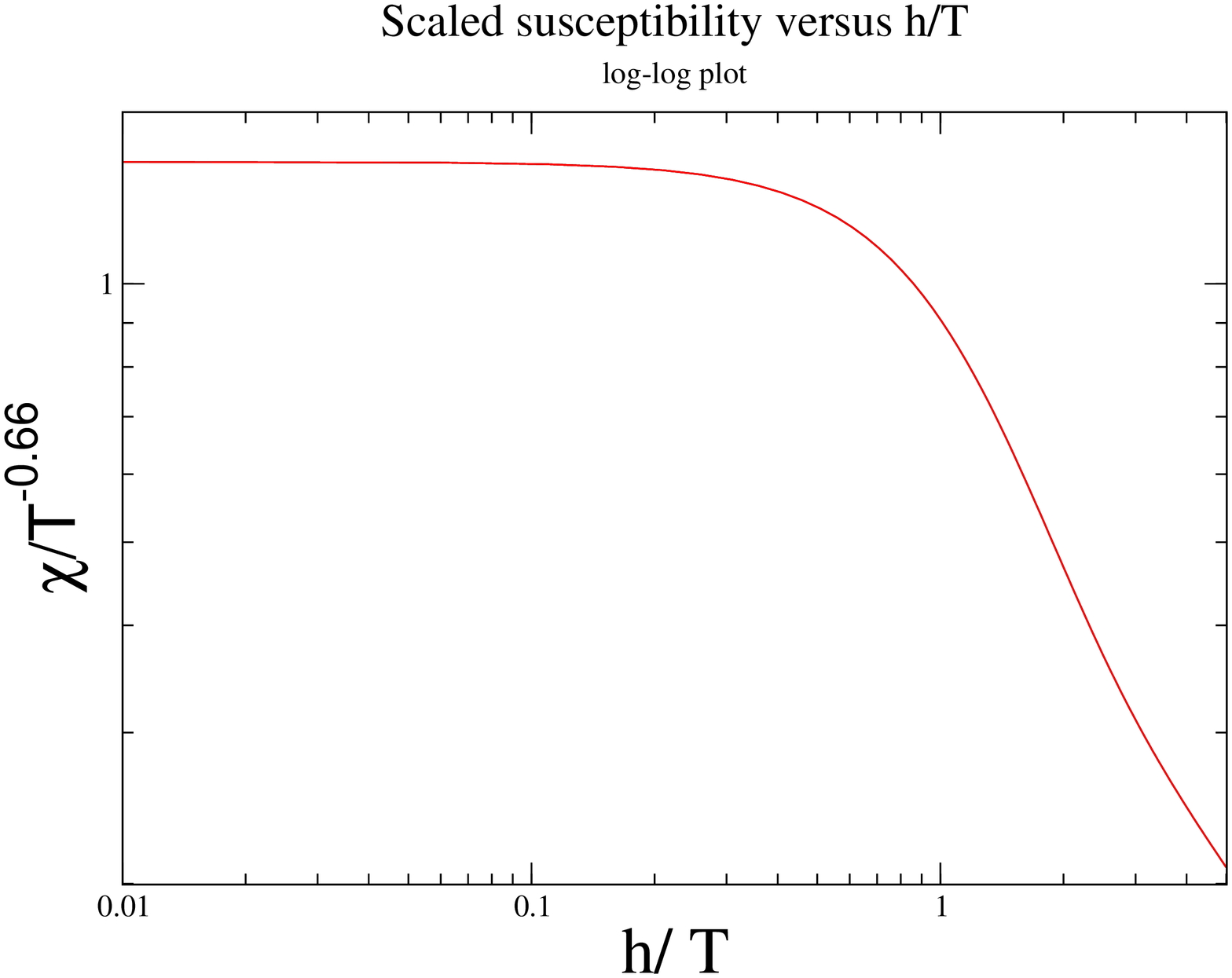}
\end{center}
\end{figure}

\begin{figure}
\begin{center}
\includegraphics[width=0.7\columnwidth,clip,angle=270]{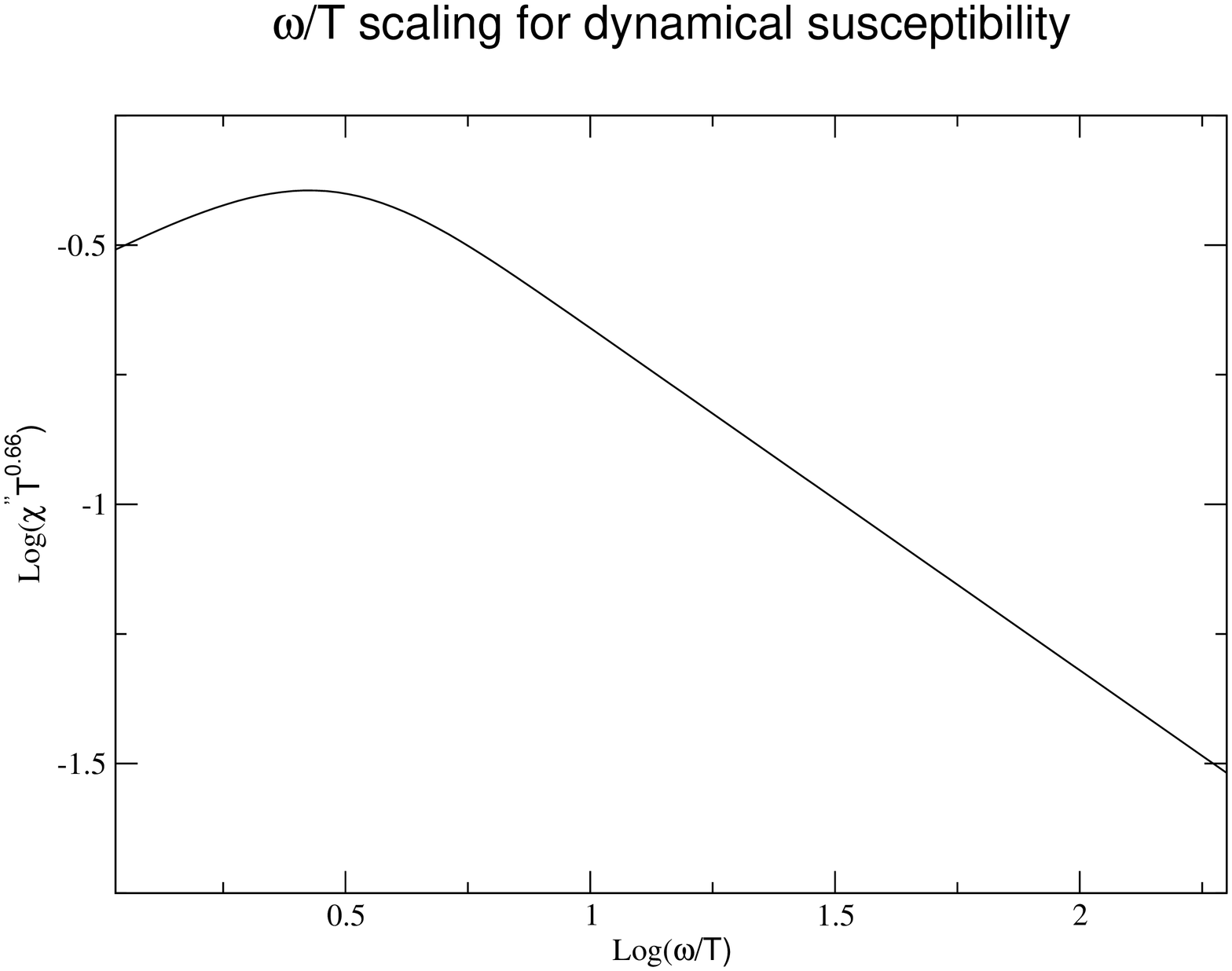}
\end{center}
\end{figure}

With this assumption, theses spins will contribute $T^{1-\alpha}$ to
the specific heat. That combined with a linear specific heat coming from the 
singlet degrees of freedom, can explain the observation of an apparent
power-law specific heat with an exponent in the range $0.5$ 
to $0.66$.\cite{helton}

Experiments show that these power law have a wide range of validity in the
Herbertsmithite materials.


\begin{thebibliography}{2}

\bibitem{pwa} P. W. Anderson, Mater. Res. Bull. 8, 153 (1973).

\bibitem{elser} C. Zeng and V. Elser, Phys. Rev. B 51, 8318 (1995).

\bibitem{mila} F. Mila, Phys. Rev. Lett. 81, 2356 (1998).

\bibitem{misguich} G. Misguich et al,
89, 137202 (2002).

\bibitem{palee} Y. Ran et al, Phys. Rev. Lett. 98, 117205 (2007).

\bibitem{mpaf}  M. Hermele et al, Phys. Rev. B 72, 104404 (2005).

\bibitem{VBC} J. B. Marston and C. Zeng, J. Appl. Phys. {\bf 69},
5962 (1991);
A. V. Syromyatnikov and S. V. Maleyev, Phys. Rev. B{\bf 66},
132408 (2002);
P. Nikolic and T. Senthil, Phys. Rev. B {\bf 68},
214415 (2003);
R. Budnik and A. Auerbach, Phys. Rev. Lett. {\bf 93},
187205 (2004);

\bibitem{ed} C. Waldtmann et al, Eur. Phys. J. B 2 501 (1998);
P. Sindzingre and C. Lhuillier, EPL 88, 27009 (2009).

\bibitem{dmrg} H. C. Jiang et al, Phys. Rev. Lett. 101, 117203 (2008).

\bibitem{singh-huse} R. R. P. Singh and D. A. Huse, Phys. Rev. B {\bf 76},
180407 (2007). 

\bibitem{ybkim} B.-J. Yang et al, Phys. Rev. B 77,
224424 (2008).

\bibitem{ms07} G. Misguich and P. Sindzingre, J. Phys.-Cond. Matt. 19, 145202 (2007).

\bibitem{laeuchli} 
A. Laeuchli, C. Lhuillier,
arXiv:0901.1065.

\bibitem{vidal}
G. Evenbly, G. Vidal,
arXiv:0904.3383.

\bibitem{tchernyshyov} Z. Hao and O. Tchernyshyov,
Phys. Rev. Lett. 103, 187203 (2009).

\bibitem{poilblanc} D. Poilblanc et al,
arXiv:0912.0724.

\bibitem{shores} M. P. Shores et al J. Am. Chem. Soc. 127, 13462 (2005).

\bibitem{helton} J. S. Helton et al, Phys. Rev. Lett. 98, 107204 (2007).

\bibitem{keren} O. Ofer et al, arXiv:cond-mat/0610540.

\bibitem{shlee} S. H. Lee et al, Nat. Mat. 6, 853 (2007).

\bibitem{olariu} A. Olariu et al, Phys. Rev. Lett. 100, 087202 (2008).

\bibitem{imai} T. Imai et al, Phys. Rev. Lett. 100, 077203 (2008).

\bibitem{devries1} M. A. De Vries et al, Phys. Rev. Lett. 100, 157205 (2008).

\bibitem{singh-huse2} R. R. P. Singh and D. A. Huse, Phys. Rev. B 77, 144415 (2008).

\bibitem{dommange} S. Dommange et al, Phys. Rev. B 68, 224416 (2003).

\bibitem{dsf} D. S. Fisher, Phys. Rev. B 50, 3799 (1994).

\bibitem{helton10} J. S. Helton et al, arXiv:1002.1091.

\bibitem{helton-pc} J. S. Helton, private communication.

\bibitem{rigol07}
M. Rigol and R. R. P. Singh, Phys. Rev. Lett. {\bf 98}, 207204
(2007).

\bibitem{sindzingre07} G. Misguich and P. Sindzingre, Eur. Phys. J. B {\bf 59},
305 (2007).

\bibitem{cepas} O. Cepas et al, Phys. Rev. B 78, 140405(R) (2008).



\bibitem{mendels} A. Zorko et al, Phys. Rev. Lett. 101, 026405 (2008).

\bibitem{mila10} I. Rousochatzakis et al, Phys. Rev. B 79, 214415 (2009).

\bibitem{devries} M. A. de Vries et al
Phys. Rev. Lett. 103, 237201 (2009).

\bibitem{kubo} T. Nakamura and K. Kubo, Phys. Rev. B 53, 6393 (1996).

\bibitem{shastry} D. Sen et al., Phys. Rev. B 53, 6401 (1996).

\bibitem{lemmens} D. Wulferding and P. Lemmens, to be published.

\bibitem{cepas2} O. Cepas et al, Phys. Rev. B 77, 
172406 (2008).

\end{thebibliography}

\end{document}